%% file: eps03.tex
\documentclass[aps,preprint,prd,tightenlines,superscriptaddress]{revtex4}
\usepackage{epsfig, amssymb}

\begin{document}
\advance\hoffset by  -4mm

\newcommand{\de}{\Delta E}
\newcommand{\mbc}{M_{bc}}
\newcommand{\bb}{B{\bar B}}
\newcommand{\qq}{q{\bar q}}
\newcommand{\ks}{K^0_S}
\newcommand{\kstar}{\bar{K}^{*0}}
\newcommand{\kpi}{K^+\pi^-}
\newcommand{\kk}{K^+ K^-}
\newcommand{\kpipin}{\kpi\pi^0}
\newcommand{\kpipipi}{\kpi\pi^-\pi^+}
\newcommand{\dkpi}{\bar{D}^0\to\kpi}
\newcommand{\dkpipin}{\bar{D}^0\to\kpipin}
\newcommand{\dkpipipi}{\bar{D}^0\to\kpipipi}
\newcommand{\phipi}{\phi\pi^+}
\newcommand{\phik}{\phi K^-}
\newcommand{\kstark}{\kstar K^+}
\newcommand{\ksk}{K_S^0 K^+}
\newcommand{\ds}{D_s}
\newcommand{\dsp}{D_s^+}
\newcommand{\dsst}{D_s^*}
\newcommand{\dsstp}{D_s^{*+}}
\newcommand{\dsj}{D_{sJ}^{*}}
\newcommand{\dsjp}{D_{sJ}^{*+}}
\newcommand{\dsphipi}{\ds\to\phipi}
\newcommand{\dskstark}{\ds\to\kstark}
\newcommand{\dsksk}{\ds\to\ksk}
\newcommand{\br}{{\cal B}}

\preprint{\vbox{ \hbox{   }
                 \hbox{BELLE-CONF-0334}
                 \hbox{EPS-ID 570}
                 \hbox{Parallel Sessions: 5, 10}
}}

\title{\quad\\[0.5cm]
  Observation of the $\dsj(2317)$ and $\dsj(2460)$ in $B$ decays}

\input{author-conf2003.tex}
\noaffiliation

\begin{abstract}
We report on a study of the $B\to D \dsj(2317)$  and $B\to D \dsj(2460)$
decays based on $123.7\times 10^6$ $\bb$ events
collected with the Belle detector at KEKB.
The $B\to D \dsj(2317)$ and $B\to D \dsj(2460)$ decays have been
observed for the first time.
We observe the $\dsj(2317)$ decay to $\ds\pi^0$ and $\dsj(2460)$ decay to
the $\dsst\pi^0$ and $\ds\gamma$ final states. We also set the 90\% CL
upper limits for the decays $\dsj(2317)\to\dsst\gamma$,
$\dsj(2460)\to\dsst\gamma$, $\dsj(2460)\to\ds\pi^0$ and
$\dsj(2460)\to\ds\pi^+\pi^-$.
\end{abstract}
\maketitle
\tighten


\section{Introduction}

Recently the BaBar collaboration reported on the observation of
a new $\ds\pi^0$ resonance with a mass of 2317 MeV and a
very narrow width~\cite{babar_dspi0}. A natural interpretation
is that this is a $p$-wave $c\bar{s}$ quark state that is below
the $D K$ threshold, which accounts for the small 
width~\cite{bardeen}. This interpretation is supported by the
observation of a $\dsst\pi^0$ resonance~\cite{footnote1}
by the CLEO collaboration~\cite{cleo_dspi0}.
Both groups observe these states in inclusive $e^+e^-$ processes.
The mass difference between the two observed states is consistent with
the expected hyperfine splitting of the $p$-wave $\ds$ 
meson doublet with total light-quark angular momentum
$j=1/2$~\cite{bardeen}. However, the masses of these states are 
considerably below potential model expectations~\cite{bartelt}, and
are nearly the same as the corresponding $c\bar{u}$
states recently measured by Belle~\cite{kuzmin}.
The low mass values have caused speculation that these
states may be more exotic than a simple $q\bar{q}$ meson 
system~\cite{cahn,lipkin,beveren,hou,fazio,godfrey}.
To clarify the nature of these states, it is necessary to
determine their quantum numbers and decay branching fractions,
particularly those for radiative decays. In this connection it is
useful to search for these states, which we refer to as
$\dsj$, in exclusive $B$ meson decay processes.
  
We search for decays of the type $B\to D \dsj$, which are expected to 
be the dominant exclusive $\dsj$ production mechanism in $B$ decays.
Because of the known properties of the parent $B$ meson, angular 
analyses of these decays could unambiguously determine the $\dsj$ 
quantum numbers. Moreover, since QCD sum rules in HQET predict
that $p$-wave mesons with $j=1/2$ should be more readily produced in $B$
decays than mesons with $j=3/2$~\cite{yao}, the observation
of $B\to D\dsj$ would provide additional support for
the $p$-wave nature of these states as well as serve as a 
check of these predictions.

In this Letter we report on a search for the $B\to D \dsj(2317)$ and
$B\to D \dsj(2460)$ decays based on a $123.7\times 10^6$ produced
$\bb$ pairs at the KEKB asymmetric energy $e^+e^-$ collider~\cite{KEKB}.
The inclusion of charge conjugate states is implicit throughout this 
report.

\section{Event selection}

The Belle detector has been described elsewhere~\cite{NIM}.
We select well measured charged tracks that have impact parameters
with respect to interaction point (IP) that are less than 0.2~cm in
the radial direction and less than 2.5~cm along the beam direction 
($z$). We also require that the transverse momentum of
the tracks be greater than 0.05~GeV$/c$ in order to reduce the 
combinatorial background from low momentum particles. 

For charged particle identification (PID), the combined information
from specific ionization in the central drift chamber ($dE/dx$),
time-of-flight scintillation counters and aerogel \v{C}erenkov
counters is used.
Charged kaons are selected with PID criteria that have
an efficiency of 88\%, a pion misidentification probability of 8\%,
and negligible contamination from protons.
All charged tracks with PID responses consistent with a pion
hypothesis that are not positively identified as electrons are 
considered as pion candidates.

Neutral kaons are reconstructed via the decay $K_S^0\to\pi^+\pi^-$
with no PID requirements for the daughter pions.
The two-pion invariant mass is required to be within 9~MeV$/c^2$
($\sim 3\sigma$) of the $K^0$ mass and the displacement of
the $\pi^+\pi^-$ vertex from the IP in the transverse
($r-\phi$) plane is required to be between 0.2~cm and 20~cm. 
The direction in the $r-\phi$ plane 
from the IP to the $\pi^+\pi^-$ vertex 
is required to agree within 0.2 radians 
with the combined momentum of the two pions.

Photon candidates are selected from energy deposit
clusters in the CsI electromagnetic
calorimeter.  Each photon candidate is required to have a laboratory 
energy greater than 30~MeV with no associated charged track, and 
a shower shape that is consistent with an electromagnetic shower.
A pair of photons with an invariant mass within 12~MeV$/c^2$ 
($\sim 2.5\sigma$) of the $\pi^0$ mass is considered as a 
$\pi^0$ candidate.

We reconstruct $\dsp$ mesons in the 
$\phipi$, $\kstark$ and $\ksk$ decay channels.
The $\phi$ mesons are reconstructed from the $\kk$ pairs with the
invariant mass within 10~MeV$/c^2$ ($\sim 2.5\Gamma$) from the
$\phi$ mass. The $\kstar$ mesons are reconstructed from $K^-\pi^+$
pairs with an invariant mass within 75~MeV$/c^2$ ($1.5\Gamma$) of the
$\kstar$ mass. After calculating the invariant mass of the
corresponding set of particles, we define a $\ds$ signal region 
as being within 12~MeV$/c^2$ ($\sim 2.5\sigma$) of the  $\ds$ 
mass. $\dsst$ mesons are reconstructed in the $\dsst\to\ds\gamma$
decay channel. The mass difference between $\dsst$ and $\ds$
candidates is required to be within 8~MeV$/c^2$ of its
nominal value ($\sim 2.5\sigma$).

We reconstruct $\bar{D}^0 (D^-)$ mesons in the $\kpi$, $\kpipipi$ and 
$\kpipin$ $(\kpi\pi^-)$ decay channels and require the invariant mass 
to be within 12~MeV$/c^2$ of the $\bar{D}^0 (D^-)$ mass.
For the $\pi^0$ from the $\dkpipin$ decay, we require that
the $\pi^0$ momentum in the $\Upsilon(4S)$ center-of-mass (CM) frame
be greater than 0.4~GeV$/c$ in order to reduce combinatorial backgrounds.

We combine $D$ candidates with  a $D_s^{(*)+}$ and 
a $\pi^0$, $\gamma$, or $\pi^+\pi^-$ pair to form $B$ mesons.
Candidate events are identified by their CM
energy difference, \mbox{$\de=(\sum_iE_i)-E_b$}, and the
beam constrained mass, $\mbc=\sqrt{E_b^2-(\sum_i\vec{p}_i)^2}$, where
$E_b$ is the beam energy and $\vec{p}_i$ and $E_i$ are the momenta and
energies of the decay products of the $B$ meson in the CM frame.
We select events with $\mbc>5.2$~GeV$/c^2$ and $|\de|<0.2$~GeV,
and define a $B$ signal region of
$5.272$~GeV$/c^2<\mbc<5.288$~GeV$/c^2$ and $|\de|<0.03$~GeV.
In the cases with more than one  candidate in an
event, the one with the $D$ and $D_s^{(*)+}$ masses
closest to the nominal values is chosen.
We use a Monte Carlo (MC) simulation to model the response of
the detector and determine the efficiency~\cite{GEANT}.

\section{Background suppression}

Variables that characterize the event topology are used
to suppress background from the
two-jet-like $e^+e^-\to\qq$ continuum process.
We require $|\cos\theta_{\rm thr}|<0.80$, where $\theta_{\rm thr}$ is 
the angle between the thrust axis of the $B$ candidate and that of the 
rest of the event; this eliminates 77\% of the continuum 
background while retaining 78\% of the signal events. 
To suppress  combinatorial background we apply a restriction on the
invariant mass of the $D$ meson and the $\pi^0$ or $\gamma$ from 
$\dsj$ decay:
$M(D\pi^0)>2.3$~GeV$/c^2$, $M(D\gamma)>2.2$~GeV$/c^2$.

\section{Results of the analysis}
\subsection{Calculation of branching fractions}

\begin{figure*}
  \includegraphics[width=0.49\textwidth] {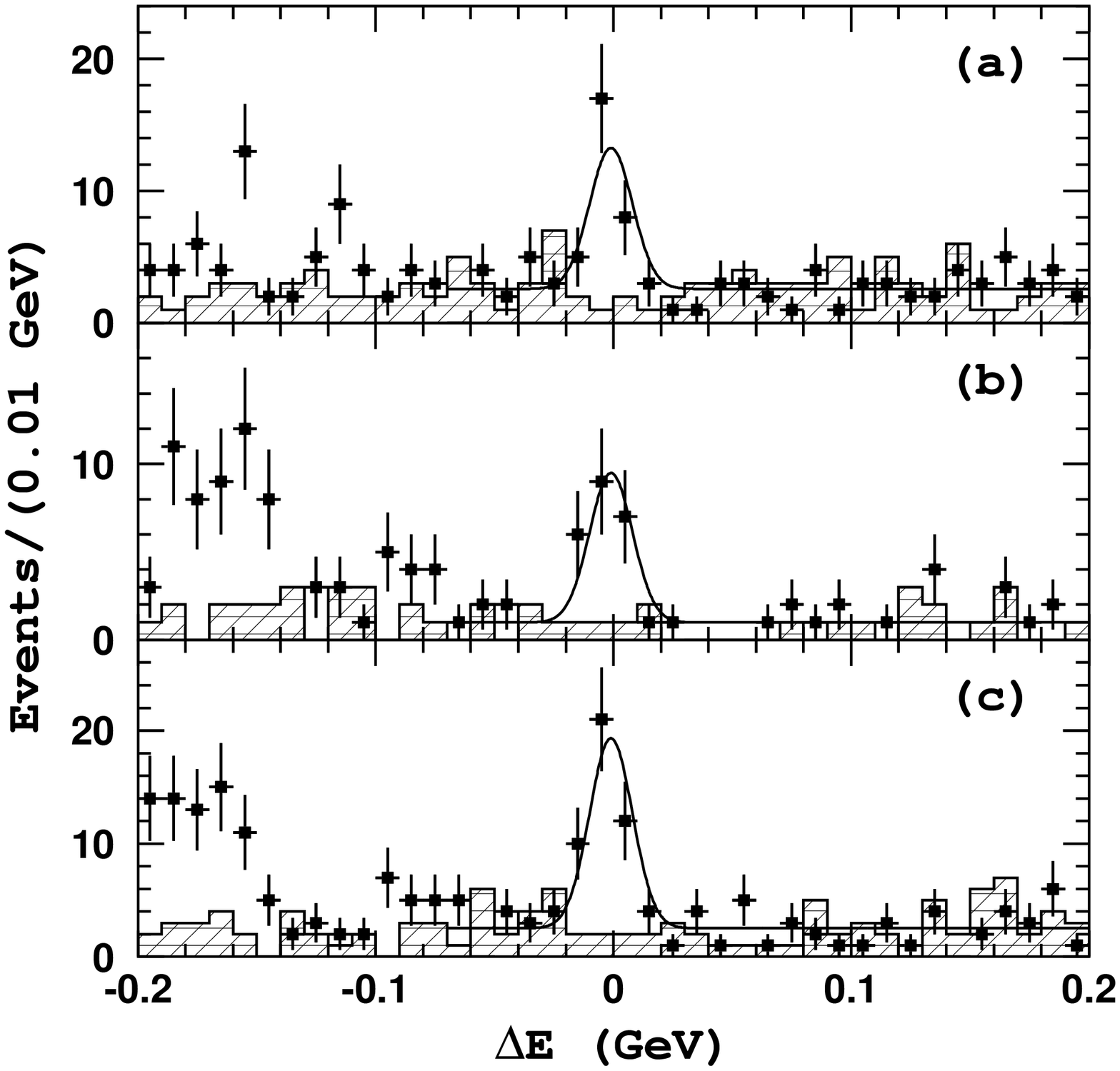} \hfill
  \includegraphics[width=0.49\textwidth] {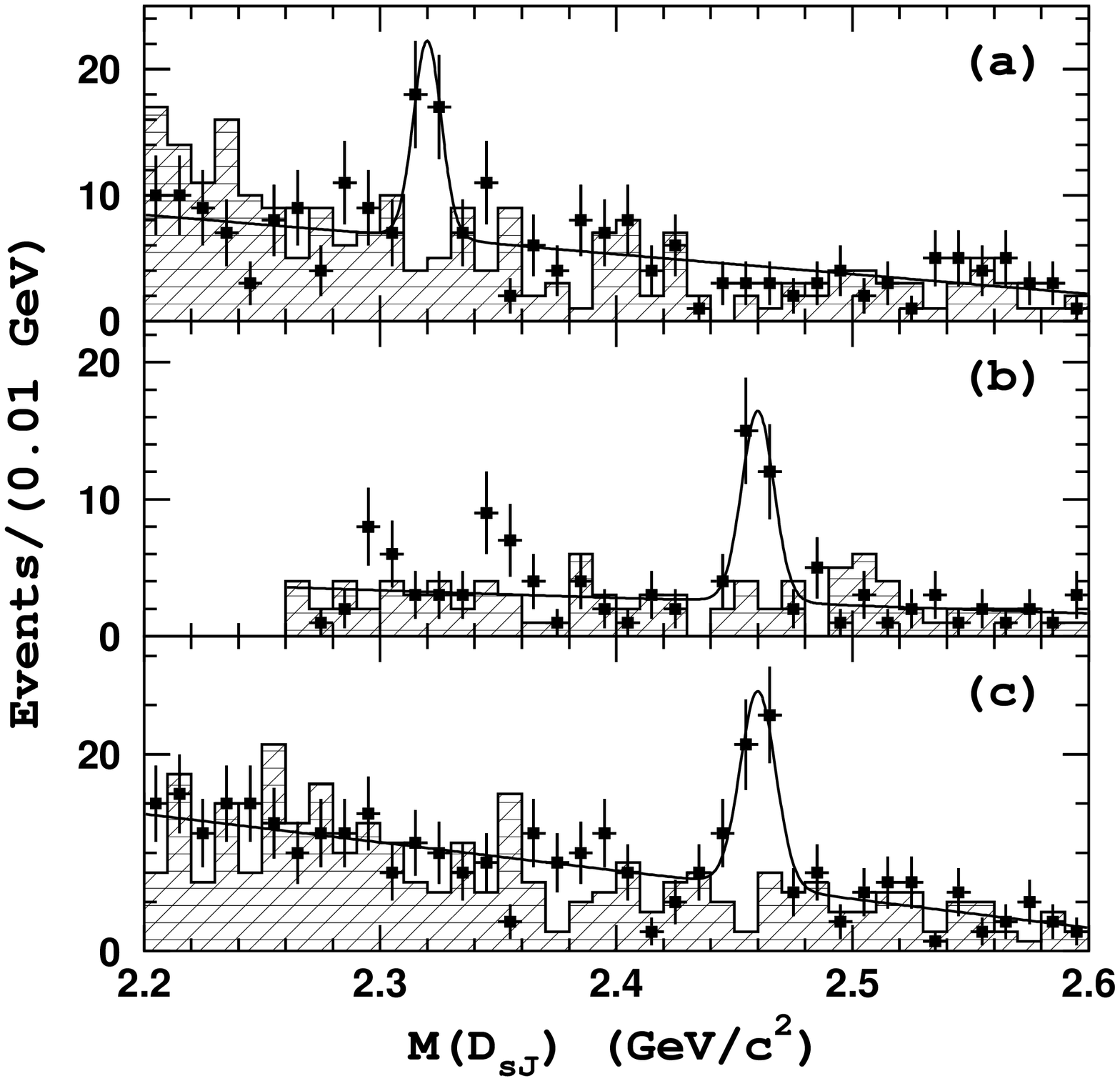}
  \caption{$\de$ (left) and $M(\dsj)$ (right) distributions for the
    $B\to D\dsj$ candidates: (a) $\dsj(2320)\to\ds\pi^0$, (b)
    $\dsj(2460)\to\dsst\pi^0$ and (c) $\dsj(2460)\to\ds\gamma$. 
    Points with errors
    represent the experimental data, hatched histograms show
    the sidebands and curves are the results of the fits.}
  \label{de_all}
\end{figure*}

The $\de$ and $\dsj$ candidate's invariant mass ($M(\dsj)$) 
distributions for $B\to D\dsj$ candidates are presented in 
Fig.~\ref{de_all}, where all $\bar{D}^0$ and $D^-$ decay modes are
combined. Each distribution is the projection of the signal region of 
the other parameter; distributions for events in the $M(\dsj)$ and 
$\de$ sidebands are shown as hatched histograms.

Clear signals are observed for the $D\dsj(2320)[\ds\pi^0]$ and 
$D\dsj(2460)[\dsst\pi^0, \ds\gamma]$ final states.
The measured masses for the $\dsj(2317)$ and $\dsj(2460)$ are
$(2319.8\pm 2.1\pm 2.0)$~MeV$/c^2$ and $(2459.2\pm 1.6\pm 2.0)$~MeV$/c^2$.

For each decay channel, the $\de$ distribution is fitted with a 
Gaussian signal and a linear background function. The Gaussian 
mean value and width are fixed to the values from a MC simulation of
signal events. The region $\de<-0.07$~GeV is excluded from the fit 
to avoid contributions from other $B$ decays of the type
$B\to D\dsj X$ where $X$ denotes an additional particle 
that is not reconstructed.
The fit results are given in Table~\ref{defit}, where the listed
efficiencies include intermediate branching fractions.  
The statistical significance of the signal quoted in Table~\ref{defit} 
is defined as
$\sqrt{-2\ln({\cal L}_0/{\cal L}_{max})}$, where ${\cal L}_{max}$ and
${\cal L}_0$ denote the maximum likelihood with the nominal
and with zero signal yield, respectively.

\begin{table*}
\caption{Branching fractions $B\to D\dsj$ decays.}
\medskip
\label{defit}
\begin{tabular*}{\textwidth}{l@{\extracolsep{\fill}}ccccc}\hline\hline
Decay channel & $\de$ yield & $M(\dsj)$ yield & Efficiency, $10^{-4}$ & 
${\cal B}$, $10^{-4}$ & Significance\\\hline
$B^+\to\bar{D}^0 \dsjp(2317)~[\dsp\pi^0]$, &
$13.7^{+5.1}_{-4.5}$ & $13.4^{+6.2}_{-5.4}$ & 1.36 & 
$8.1^{+3.0}_{-2.7}\pm 2.4$ & $5.0\sigma$\\
$B^0\to D^- \dsjp(2317)~[\dsp\pi^0]$ &
$10.3^{+3.9}_{-3.1}$ & $10.8^{+4.2}_{-3.6}$ & 0.97 &
$8.6^{+3.3}_{-2.6}\pm 2.6$ & $6.1\sigma$\\

$B^+\to\bar{D}^0 \dsjp(2317)~[\dsstp\gamma]$ &
$3.4^{+2.8}_{-2.2}$ & $2.1^{+4.1}_{-3.4}$ & 1.15 & 
$2.4^{+2.0}_{-1.5}(<5.7)$ & ---\\
$B^0\to D^- \dsjp(2317)~[\dsstp\gamma]$ &
$2.3^{+2.5}_{-1.9}$ & $1.6^{+2.4}_{-1.9}$ & 0.71 & 
$2.6^{+2.8}_{-2.2}(<7.1)$ & ---\\

$B^+\to\bar{D}^0 \dsjp(2460)~[\dsstp\pi^0]$ &
$7.2^{+3.7}_{-3.0}$ & $8.9^{+4.0}_{-3.3}$ & 0.49 & 
$11.9^{+6.1}_{-4.9}\pm 3.6$ & $2.9\sigma$\\
$B^0\to D^- \dsjp(2460)~[\dsstp\pi^0]$ &
$11.8^{+3.8}_{-3.2}$ & $14.9^{+4.4}_{-3.9}$ & 0.42 & 
$22.7^{+7.3}_{-6.2}\pm 6.8$ & $6.5\sigma$\\

$B^+\to\bar{D}^0 \dsjp(2460)~[\dsp\gamma]$ &
$19.1^{+5.6}_{-5.0}$ & $20.2^{+7.2}_{-6.9}$ & 2.75 & 
$5.6^{+1.6}_{-1.5}\pm 1.7$ & $5.0\sigma$\\
$B^0\to D^- \dsjp(2460)~[\dsp\gamma]$ &
$18.5^{+5.0}_{-4.3}$ & $19.6^{+5.6}_{-4.9}$ & 1.83 & 
$8.2^{+2.2}_{-1.9}\pm 2.5$ & $6.5\sigma$\\

$B^+\to\bar{D}^0 \dsjp(2460)~[\dsstp\gamma]$ &
$4.4^{+3.8}_{-3.3}$ & $8.2^{+4.0}_{-3.4}$ & 1.15 & 
$3.1^{+2.7}_{-2.3}(<7.5)$ & ---\\
$B^0\to D^- \dsjp(2460)~[\dsstp\gamma]$ &
$1.1^{+1.8}_{-1.2}$ & $0.2^{+1.8}_{-1.2}$ & 0.71 & 
$1.3^{+2.0}_{-1.4}(<4.6)$ & ---\\

$B^+\to\bar{D}^0 \dsjp(2460)~[\dsp\pi^+\pi^-]$ &
$<4.0$ & $-2.2^{+2.0}_{-1.6}$ & 1.89 & $<1.7$ & ---\\
$B^0\to D^- \dsjp(2460)~[\dsp\pi^+\pi^-]$ & 
$<2.5$ & $-1.2^{+2.7}_{-2.0}$ & 1.35 & $<1.5$ & ---\\

$B^+\to\bar{D}^0 \dsjp(2460)~[\dsp\pi^0]$ & $<2.4$ & 
$1.0^{+2.7}_{-2.0}$ & 0.94 & $<2.1$ & ---\\
$B^0\to D^- \dsjp(2460)~[\dsp\pi^0]$ & $<2.4$ & 
$0.3^{+1.8}_{-1.2}$ & 0.68 & $<2.8$ & ---\\
\hline\hline
\end{tabular*}
\end{table*}

The combined fit quoted in Table~\ref{simulfit} is the summed results
for the $\bar{D}^0$ and $D^-$ modes assuming isospin invariance.
The normalization of the background in each sub-mode is allowed to 
float while the signal yields are required to satisfy the constraint
$N_i= N_{\bb}\cdot{\cal B}(B\to D\dsj)\cdot\varepsilon_i\, ,$
where the branching fraction ${\cal B}(B\to D\dsj)$ is a fit parameter;
$N_{\bb}$ is the number of $\bb$ pairs and $\varepsilon_i$ is the 
efficiency, which includes all intermediate branching fractions.
From the two $B\to D\dsj(2460)$ branching fraction measurements, we 
determine the ratio $\br(\dsj(2460)\to\ds\gamma)/
\br(\dsj(2460)\to\dsst\pi^0)=0.38\pm0.11\pm 0.04$.

\begin{table*}
\caption{Combined fit results.}
\medskip
\label{simulfit}
\begin{tabular*}{\textwidth}{l@{\extracolsep{\fill}}cc}\hline\hline
Decay channel & ${\cal B}$, $10^{-4}$ & Significance\\\hline
$B\to D \dsj(2317)~[\ds\pi^0]$ & 
$8.5^{+2.1}_{-1.9}\pm 2.6$ & $6.1\sigma$\\
$B\to D \dsj(2317)~[\dsst\gamma]$ & 
$2.5^{+2.0}_{-1.8}(<5.8)$ & $1.8\sigma$\\

$B\to D\dsj(2460)~[\dsst\pi^0]$ & 
$17.8^{+4.5}_{-3.9}\pm 5.3$ & $6.4\sigma$\\

$B\to D\dsj(2460)~[\ds\gamma]$ &
$6.7^{+1.3}_{-1.2}\pm 2.0$ & $7.4\sigma$\\

$B\to D\dsj(2460)~[\dsst\gamma]$ &
$2.7^{+1.8}_{-1.5}(<5.6)$ & $2.1\sigma$\\

$B\to D\dsj(2460)~[\ds\pi^+\pi^-]$ & $<1.2$ & ---\\

$B\to D\dsj(2460)~[\ds\pi^0]$ & $<1.4$ & ---\\\hline\hline
\end{tabular*}
\end{table*}

\begin{figure}
  \includegraphics[width=0.45\textwidth] {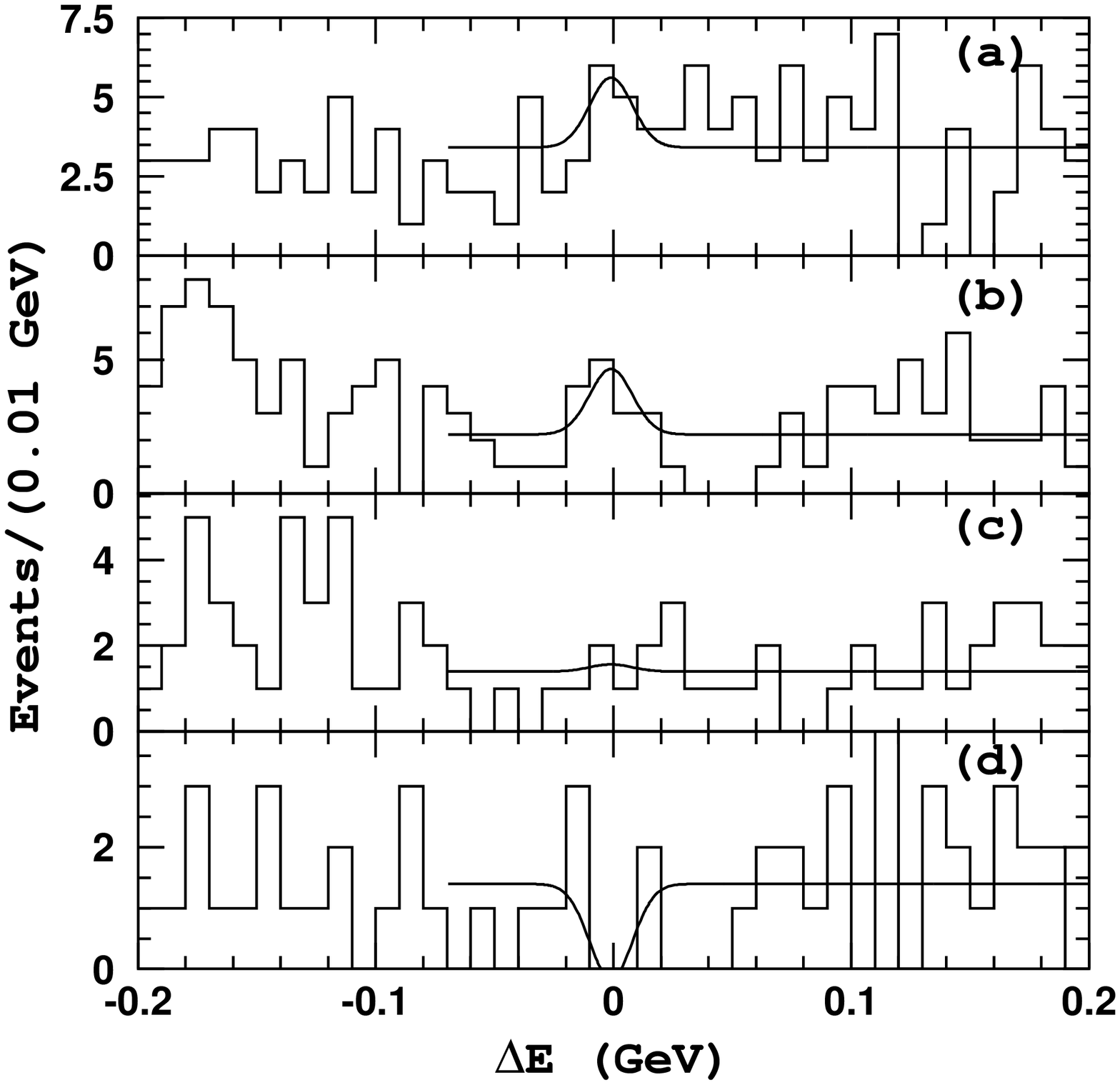}
  \caption{$\de$ distributions for the decay channels with not
    significant signals: (a) $D\dsj(2320)[\dsst\gamma]$, (b)
    $D\dsj(2460)[\dsst\gamma]$, (c) $D\dsj(2460)[\ds\pi^+\pi^-]$, 
    (d) $D\dsj(2460)[\ds\pi^0]$.
    Open histograms represent the experimental data
    and curves show the results of the fits.}
  \label{de_ul}
\end{figure}

The signals for the $B\to D \dsj(2317)[\ds\pi^0]$ and
$B\to D \dsj(2460)[\dsst\pi^0, \ds\gamma]$ channels have 
greater than $5\sigma$ statistical significance.
Figure~\ref{de_ul} shows the $\de$ distributions for the
other channels, where significant signals are not seen. 
We set 90\% confidence level (CL) upper limits for 
these modes based on an event yield
$N$ that is calculated from the relation 
$\int^N_0 {\cal L}(n) dn=0.9\int^{\infty}_0 {\cal L}(n) dn$, 
where ${\cal L}(n)$ is the maximum likelihood with the signal 
yield equal to $n$. 

\subsection{Angular analysis}
\begin{figure}
  \includegraphics[width=0.4\textwidth] {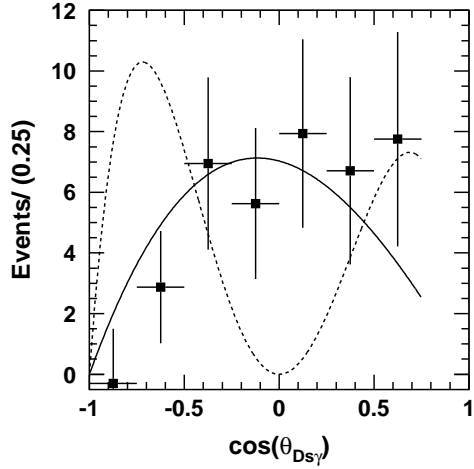}
  \caption{The $\dsj(2460)\to\ds\gamma$ helicity distribution.
   The points with error bars are the results of fits to the $\de$ 
   spectra for experimental events in the corresponding bin. 
   Solid and dashed curves are MC predictions for a $J=1$ and $J=2$
   hypothesis, respectively. The highest bin has 
   no events because of cut to the $D\gamma$ invariant mass.} 
  \label{check}
\end{figure}

The helicity angle analysis can provide information about spin of the
decaying particle. We study the helicity angle distribution for the
$\dsj(2460)\to\ds\gamma$ decay.

The helicity angle $\theta_{\ds\gamma}$ is defined as the angle
between the $\dsj(2460)$ momentum in the $B$ meson rest frame and the 
$\ds$ momentum in the $\dsj(2460)$ rest frame.
The distribution shown in Fig.~\ref{check} is consistent with MC 
expectations for a $J=1$ hypothesis for the $\dsj(2460)$ 
($\chi^2/$n.d.f$=5/6$), and contradicts the $J=2$ hypothesis
($\chi^2/$n.d.f$=44/6$).

\subsection{Cross checks \& systematic uncertainties}

We study the possible feed-across between all studied $\dsj$ decay
modes using MC. We also analyse a MC sample of generic 
$\bb$ events corresponding to our data sample.
No peaking background is found.

As a check, we apply a similar procedure to decay chains with the 
same final states: $B\to D\dsst$, $B\to D^*\ds$ and $B\to D^*\dsst$.
For each mode, we measure branching fractions that 
are consistent with the world average values~\cite{PDG}.

The following sources of systematic errors are considered:
tracking efficiency (1-2\% per track), kaon identification
efficiency (1\%), $\pi^0$ efficiency (6\%), $K^0_S$ reconstruction 
efficiency (6\%), efficiency for slow pions 
from $D^*\to D\pi$ decays (8\%), $D$ branching fraction
uncertainties (2\%-6\%), signal and background shape
parameterization (4\%) and MC statistics (3\%).
The uncertainty in the tracking efficiency is estimated using 
partially reconstructed $D^{*+}\to D^0\pi^+$, $D^0\to K_S^0\pi^+\pi^-$ 
decays. The kaon identification uncertainty is determined 
from $D^{*+}\to D^0\pi^+$, $D^0\to K^-\pi^+$ decays.
The $\pi^0$ reconstruction uncertainty is obtained using 
$\bar{D}^0$ decays to $\kpi$ and $\kpipin$.
We assume equal production rates for $B^+B^-$ and $B^0\bar B^0$ pairs 
and do not include the uncertainty related to this assumption in the 
total systematic error. For the calculation of the branching
fractions, the errors in the $\ds$ meson branching fractions are
taken into account. These uncertainties are dominated by the error of
the $\dsphipi$ branching ratio of 25\%~\cite{PDG}. The overall
systematic uncertainty is 30\%.

\section{Conclusion}

In summary, we report the first observation of $B\to D\dsj(2320)$ 
and $B\to D\dsj(2460)$ decays. 
The measured branching fractions with the corresponding statistical
significances are presented in Table~\ref{simulfit}.
The angular analysis of the $\dsj(2460)\to\ds\gamma$ decay supports
the hypothesis that $\dsj(2460)$ is a $1^+$ state.

We wish to thank the KEKB accelerator group for the excellent
operation of the KEKB accelerator.
We acknowledge support from the Ministry of Education,
Culture, Sports, Science, and Technology of Japan
and the Japan Society for the Promotion of Science;
the Australian Research Council
and the Australian Department of Education, Science and Training;
the National Science Foundation of China under contract No.~10175071;
the Department of Science and Technology of India;
the BK21 program of the Ministry of Education of Korea
and the CHEP SRC program of the Korea Science and Engineering Foundation;
the Polish State Committee for Scientific Research
under contract No.~2P03B 01324;
the Ministry of Science and Technology of the Russian Federation;
the Ministry of Education, Science and Sport of the Republic of Slovenia;
the National Science Council and the Ministry of Education of Taiwan;
and the U.S.\ Department of Energy.

\end{document}

%% file: author-conf2003.tex
\affiliation{Aomori University, Aomori}
\affiliation{Budker Institute of Nuclear Physics, Novosibirsk}
\affiliation{Chiba University, Chiba}
\affiliation{Chuo University, Tokyo}
\affiliation{University of Cincinnati, Cincinnati, Ohio 45221}
\affiliation{University of Frankfurt, Frankfurt}
\affiliation{Gyeongsang National University, Chinju}
\affiliation{University of Hawaii, Honolulu, Hawaii 96822}
\affiliation{High Energy Accelerator Research Organization (KEK), Tsukuba}
\affiliation{Hiroshima Institute of Technology, Hiroshima}
\affiliation{Institute of High Energy Physics, Chinese Academy of Sciences, Beijing}
\affiliation{Institute of High Energy Physics, Vienna}
\affiliation{Institute for Theoretical and Experimental Physics, Moscow}
\affiliation{J. Stefan Institute, Ljubljana}
\affiliation{Kanagawa University, Yokohama}
\affiliation{Korea University, Seoul}
\affiliation{Kyoto University, Kyoto}
\affiliation{Kyungpook National University, Taegu}
\affiliation{Institut de Physique des Hautes \'Energies, Universit\'e de Lausanne, Lausanne}
\affiliation{University of Ljubljana, Ljubljana}
\affiliation{University of Maribor, Maribor}
\affiliation{University of Melbourne, Victoria}
\affiliation{Nagoya University, Nagoya}
\affiliation{Nara Women's University, Nara}
\affiliation{National Kaohsiung Normal University, Kaohsiung}
\affiliation{National Lien-Ho Institute of Technology, Miao Li}
\affiliation{Department of Physics, National Taiwan University, Taipei}
\affiliation{H. Niewodniczanski Institute of Nuclear Physics, Krakow}
\affiliation{Nihon Dental College, Niigata}
\affiliation{Niigata University, Niigata}
\affiliation{Osaka City University, Osaka}
\affiliation{Osaka University, Osaka}
\affiliation{Panjab University, Chandigarh}
\affiliation{Peking University, Beijing}
\affiliation{Princeton University, Princeton, New Jersey 08545}
\affiliation{RIKEN BNL Research Center, Upton, New York 11973}
\affiliation{Saga University, Saga}
\affiliation{University of Science and Technology of China, Hefei}
\affiliation{Seoul National University, Seoul}
\affiliation{Sungkyunkwan University, Suwon}
\affiliation{University of Sydney, Sydney NSW}
\affiliation{Tata Institute of Fundamental Research, Bombay}
\affiliation{Toho University, Funabashi}
\affiliation{Tohoku Gakuin University, Tagajo}
\affiliation{Tohoku University, Sendai}
\affiliation{Department of Physics, University of Tokyo, Tokyo}
\affiliation{Tokyo Institute of Technology, Tokyo}
\affiliation{Tokyo Metropolitan University, Tokyo}
\affiliation{Tokyo University of Agriculture and Technology, Tokyo}
\affiliation{Toyama National College of Maritime Technology, Toyama}
\affiliation{University of Tsukuba, Tsukuba}
\affiliation{Utkal University, Bhubaneswer}
\affiliation{Virginia Polytechnic Institute and State University, Blacksburg, Virginia 24061}
\affiliation{Yokkaichi University, Yokkaichi}
\affiliation{Yonsei University, Seoul}
  \author{K.~Abe}\affiliation{High Energy Accelerator Research Organization (KEK), Tsukuba} 
  \author{K.~Abe}\affiliation{Tohoku Gakuin University, Tagajo} 
  \author{N.~Abe}\affiliation{Tokyo Institute of Technology, Tokyo} 
  \author{R.~Abe}\affiliation{Niigata University, Niigata} 
  \author{T.~Abe}\affiliation{High Energy Accelerator Research Organization (KEK), Tsukuba} 
  \author{I.~Adachi}\affiliation{High Energy Accelerator Research Organization (KEK), Tsukuba} 
  \author{Byoung~Sup~Ahn}\affiliation{Korea University, Seoul} 
  \author{H.~Aihara}\affiliation{Department of Physics, University of Tokyo, Tokyo} 
  \author{M.~Akatsu}\affiliation{Nagoya University, Nagoya} 
  \author{M.~Asai}\affiliation{Hiroshima Institute of Technology, Hiroshima} 
% ...
\collaboration{The Belle Collaboration}

%% file: eps03.bbl
\begin{thebibliography}{99}
\bibitem{footnote1}{In the heavy $c$-quark mass limit, one expects two 
  doublets of $c\bar{s}$ states with quantum 
  numbers $J^P=0^+$, $1^+$ and $1^+$, $2^+$. The second doublet has 
  been observed in $D^{(*)}K$ decays.}
\end{thebibliography}
